\documentclass[twocolumn,superscriptaddress,showpacs]{revtex4}

\usepackage{amsmath,amssymb,graphicx,color,times,psfrag}

\renewcommand\epsilon{\varepsilon}
\renewcommand\phi{\varphi}
\renewcommand\vec[1]{\boldsymbol{\mathrm{#1}}}
\newcommand\unitvec[1]{\vec{\hat {#1}}}
\newcommand\dotprod{\boldsymbol{\cdot}}

\newcommand\expect[1]{\left\langle\vphantom{\big(}#1\right\rangle}

\newcommand{\eq}[1]{Eq.~\eqref{eq:#1}}
\newcommand{\fig}[1]{Fig.~\ref{fig:#1}}
\renewcommand{\sec}[1]{Section~\ref{sec:#1}}     

\newcommand\df{{d_\text{f}}}
\newcommand\dw{{d_\text{w}}}

\newcommand\tX{t_\mathsf{x}}
\newcommand\tauB{\tau_\text{B}}
\newcommand\kB{k_\text{B}}

\newcommand\ie{i.\,e., }
\newcommand\eg{e.\,g., }

\begin{document}

\newlength{\figwidth}
\setlength{\figwidth}{\linewidth}

\title{Critical dynamics of ballistic and Brownian particles in a heterogeneous environment}
\author{Felix H{\"o}f\/ling}
\affiliation{Arnold Sommerfeld Center for Theoretical Physics (ASC)  and Center for
NanoScience (CeNS), Fakult{\"a}t f{\"u}r Physik,
Ludwig-Maximilians-Universit{\"a}t M{\"u}nchen, Theresienstra{\ss}e 37,
80333 M{\"u}nchen, Germany}
\affiliation{Hahn-Meitner-Institut Berlin,
Abteilung Theorie, Glienicker Stra{\ss}e 100, 14109 Berlin, Germany}
\author{Tobias Munk}
\affiliation{Arnold Sommerfeld Center for Theoretical Physics (ASC)  and Center for
NanoScience (CeNS), Fakult{\"a}t f{\"u}r Physik,
Ludwig-Maximilians-Universit{\"a}t M{\"u}nchen, Theresienstra{\ss}e 37,
80333 M{\"u}nchen, Germany}
\author{Erwin Frey}
\affiliation{Arnold Sommerfeld Center for Theoretical Physics (ASC)  and Center for
NanoScience (CeNS), Fakult{\"a}t f{\"u}r Physik,
Ludwig-Maximilians-Universit{\"a}t M{\"u}nchen, Theresienstra{\ss}e 37,
80333 M{\"u}nchen, Germany}
\author{Thomas Franosch}
\affiliation{Arnold Sommerfeld Center for Theoretical Physics (ASC)  and Center for
NanoScience (CeNS), Fakult{\"a}t f{\"u}r Physik,
Ludwig-Maximilians-Universit{\"a}t M{\"u}nchen, Theresienstra{\ss}e 37,
80333 M{\"u}nchen, Germany}

\begin{abstract}
The dynamic properties of a classical tracer particle in a random, disordered medium are investigated
close to the localization transition.
For Lorentz models obeying Newtonian and diffusive motion at the microscale,
we have performed large-scale computer simulations, demonstrating that universality holds at long times in the immediate vicinity of the transition. The scaling function describing the crossover from  anomalous transport to diffusive motion is found to vary extremely slowly and spans at least 5 decades in time.
To extract the scaling function, one has to allow for the leading universal corrections to scaling.
Our findings suggest that  apparent power laws with varying exponents generically occur and dominate experimentally accessible time windows as soon as the heterogeneities cover a decade in length scale.
We extract the divergent length scales, quantify the spatial heterogeneities in terms of the non-Gaussian parameter, and corroborate our results by a thorough finite-size analysis.
\end{abstract}


\maketitle


\section{Introduction}
\label{sec:intro}

Heterogeneous materials abound in synthetic products and in nature; they are composed of domains of different
materials or phases, with characteristic dimensions covering a wide range
of length scales.
A physical understanding of their macroscopic
properties, such as mechanical elasticity, electrical conductivity, particle transport,
or fluid permeability has far reaching consequences for applications in
material science, nano-chemistry, oil recovery, and even biology.
Examples include
anomalous transport of tracers in porous soil columns~\cite{Cortis:2004},
slow diffusion of sodium ions in sodium silicates~\cite{Meyer:2004,Voigtmann:2006},
and transport in colloidal gels close to gelation~\cite{Romer:2000,Pham:2002,DelGado:2000,Zaccarelli:2005,Abete:2007}
In biology, the dense packing of differently sized proteins, lipids, and sugars in the cell cytoplasm is summarized as  macromolecular crowding~\cite{Ellis:2001a,Ellis:2001,Hall:2003}. It leads to a suppression of diffusion with increasing molecular weight~\cite{Arrio-Dupont:2000} and to anomalous diffusion~\cite{Saxton:1994}, observed in eukaryotes~\cite{Luby-Phelps:1987,Caspi:2002,Tolic-Norrelykke:2004,Weiss:2004,Guigas:2007} and bacteria~\cite{Golding:2006}.

Transport of tagged ions, macromolecules, or nanoparticles in such heterogeneous environments is strongly hindered,
since the presence of a variety of components reduces the accessible volume to a small fraction of three-dimensional space.
In computer simulations, one observes a drastic suppression of the diffusion coefficient by up to several orders of magnitude upon decreasing the porosity, \ie the fraction of the non-excluded volume. There, the medium is often modeled by randomly placed obstacles~\cite{Torquato:1989,Kim:1992,Viramontes-Gamboa:1995}, but recent studies investigate also realistically reconstructed media, \eg Vycor glass~\cite{Kainourgiakis:1999} and North Sea chalk~\cite{Kainourgiakis:2005}.

In the above examples, three major transport phenomena are observed: normal diffusion, immobilization or localization, and anomalous transport.
We will demonstrate in the following that all three aspects may be unified into the concept of transport in a disordered, heterogeneous medium with a percolation transition; such a transition entails a critical point with a divergent correlation length.
The asymptotic behavior of this length scale in the critical region, together with the intrinsic properties at criticality, is encoded in the renormalization group flow; therefrom, all macroscopic observables (such as the diffusion coefficient) can be inferred in principle. This leads to the postulate of universality: systems sharing the same critical point exhibit the same universal scaling laws in the critical regime. Consequently, one expects a generic mechanism for slow, anomalous transport in a heterogeneous environment.

In a recent study on the Lorentz model, \ie for a ballistic tracer in a porous medium, we have shown that a continuum percolation transition of the accessible volume is responsible for the suppression of the diffusion coefficient~\cite{Lorentz_PRL:2006}. We have successfully applied the theory of critical dynamic scaling to explain the dynamics over many decades of time and length scales, covering a large range of porosities. This analysis will be extended and further substantiated here.
In addition, we present simulation results for Brownian particles in a porous medium, modeling macroscopic particles, \eg proteins, that experience thermal fluctuations from a solvent. Universality implies that the long-time behavior of dynamic observables close to the transition is independent of the microscopic details. It predicts that the critical exponents of the anomalous diffusion and the scaling of the diffusion coefficient are the same and moreover that the scaling functions coincide. The equivalence of ballistic and Brownian particles is well known for molecular and colloidal glasses~\cite{Gleim:1998}, and it is expected to hold generally for slow dynamics. By a direct comparison of ballistic and Brownian particles, we test this hypothesis and give an estimate of the asymptotic scaling regime. The latter is essential for the interpretation of experiments, since it allows to assess the applicability of the asymptotic laws to a specific measurement window.

Let us summarize the different transport phenomena briefly. Diffusion is well known and understood since more than a century now~\cite{Haenggi:2005,Frey:2005}. It is connected with Brownian motion, observed for a large particle kicked around by the surrounding gas or solvent molecules. The corresponding mean-square displacement grows linearly in time,
\begin{equation}
\delta r^2(t):=\expect{\Delta \vec R(t)^2}=2d\,D t,
\end{equation}
where $\Delta \vec R(t)$ denotes the displacement of the particle after a sufficiently long time lag
$t$, $d$ is the space dimension, and $\expect{\dots}$ an appropriate average.
The phenomenon of localization is reflected in a plateau in the mean-square displacement,
defining the localization length~$\ell$,
\begin{equation}
\delta r^2(t)=\textit{const}=\ell^2.
\end{equation}
Particles may be immobilized due to chemical binding, or they get trapped in cages formed by the surroundings.

Anomalous transport is certainly the most fascinating phenomenon among the three. Contrary to normal diffusion, it is not simply a consequence of the central limit theorem; in particular, it requires the non-trivial presence of either a broad distribution of time scales or long-ranged correlations~\cite{Bouchaud:1990}. The mean-square displacement exhibits a power law growth with a fractal exponent,
\begin{equation}
\delta r^2(t)\sim t^{2/\dw};
\end{equation}
here we shall discuss subdiffusion, $\dw>2$.
In most cases, it originates in one of two mechanisms: binding to finite traps with a broad distribution of binding rates, or confined motion in a spatially non-uniform, heterogeneous medium.
Both scenarios may be easily distinguished experimentally, since binding rates obey an Arrhenius law in general, and thus, the dynamic coefficients are very sensitive to temperature changes.
Steric interaction in contrast is insensitive to temperature and often conveniently modeled by hard potentials.

The article is organized as follows: In the first part, we introduce the Lorentz model (\sec{Lorentz_model}), outline its relation to continuum percolation, and review some theoretical results relevant for the subsequent analysis (\sec{percolation}).
In \sec{simulations}, we set forth the applied simulation techniques, while \sec{simulation_results} is devoted to the scaling behavior of diffusion coefficients, length scales, the mean-square displacement, and the non-Gaussian parameter, complemented by a finite-size scaling analysis.

\section{The Lorentz model}
\label{sec:Lorentz_model}

The Lorentz model is a generic model covering all of the above aspects of transport in a heterogeneous environment.
It has attracted the attention of researchers in statistical physics over more than a century by now and was applied to various contexts. Lorentz introduced the model in 1905 as a microscopic justification of the Drude conductivity of a metal~\cite{Lorentz:1905}: a non-interacting electron gas is scattered off randomly distributed ions. Equivalently, one may think of a single ballistic and point-like tracer particle with velocity $v=|\vec v|$, exploring the void space between randomly, uniformly, and independently placed scatterers of number density~$n$. A hard-core interaction with the scatterers is usually employed, yielding impenetrable obstacles of radius~$\sigma$. This interaction potential is equivalent to a tracer particle and obstacles sharing the same radius $\sigma/2$. The kinetic energy of the particle is conserved, and the unit time and length scales are set by $\sigma$ and $t_0:=v^{-1}\sigma$, respectively. The only control parameter of the model is the dimensionless obstacle density $n^*:=n\sigma^d$; it is directly linked with the porosity of the medium,
\begin{equation}
\phi=\exp(-4\pi n^*/3).
\end{equation}
Statistical averages include different initial positions (restricted to the void space) and velocities (with fixed magnitude) and different realizations of the disorder; in particular, the tracer particle is not restricted to the percolating cluster.

Although the model is a deterministic system, the initial velocity
of the particle is quickly randomized by subsequent collisions with the
obstacles, yielding diffusive motion on large time scales.
Such a stochastic trajectory may be thought of as a hopping process with a general distribution of waiting times $p_\tau(\tau)$ and displacements $p_{\vec a}(\vec a)$~\cite{Bouchaud:1990}.
Provided that the first and second moments exist, uncorrelated steps generate diffusive motion on large time scales with diffusion coefficient
\begin{equation}
\label{eq:D_random_walk}
D=\frac{\expect{\vec a^2}}{2d \expect{\tau}}=\frac{v^2\expect{\tau^2}}{2d \expect \tau};
\end{equation}
the second form refers to a ballistic particle with fixed kinetic energy.
In the Lorentz model, the collisions constitute a Poisson process with average rate $\tau_c^{-1}$, thus the waiting time distribution has first and second moments $\expect{\tau}=\tau_c$ and $\expect{\tau^2}=2\tau_c^2$, respectively.
The mean collision rate follows from the condition to encounter no obstacles in a corridor of volume $\pi\sigma^2 (v \tau_c)=1/n$ in $d=3$ dimensions, and thus $\tau_c^{-1}=\pi n^* v/\sigma$.
Since the differential scattering cross-section for a sphere is isotropic, it corresponds to the transport cross-section describing the transfer of momentum; subsequent collisions are uncorrelated in the dilute limit, $n^*\to 0$.
Under these conditions, \eq{D_random_walk} yields the correct diffusion coefficient~\cite{Lorentz:1905},
\begin{equation}
  \label{eq:D_Boltzmann}
  D_0=\frac{v \sigma}{3 \pi n^*} \quad \text{for} \quad n^*\to 0,
\end{equation}
recovering the leading order in an asymptotic low-density expansion~\cite{vanLeeuwen:1967,Weijland:1968}.

At finite densities, spatial correlations between obstacles induce persistent anti-correlations of the velocity, which are reflected in a negative tail of the velocity auto-correlation function, $\psi(t)=\expect{\vec v(t)\dotprod\vec v(0)}/v^2\sim -t^{-d/2-1}$ for $t\to\infty$~\cite{Ernst:1971a,vanBeijeren:1982}. Since the integral over $\psi(t)$ is related to the diffusion coefficient, such a tail reduces $D$.
The long-time tail can be thought of as a consequence of backscattering events preferring the return of the particle to its origin, emphasizing the importance of the topology of the obstacle matrix.
The exponent of the tail is universal, \ie it does not depend on the density of the scatterers. But as the density is increased,
a pre-asymptotic, negative tail emerges, which suppresses the diffusion coefficient additionally~\cite{Goetze:1981a,Goetze:1981b,Lorentz_LTT:2007}. At a critical obstacle density $n^*_c$, the pre-asymptotic tail persists for all times, yielding exactly zero diffusivity.

At higher densities, diffusion is absent too: the tagged particle is trapped,
and the mean-square displacement saturates. The transition from diffusion to
localization has the signatures of a continuous phase transition; it exhibits
power-law divergences of physical quantities with universal critical exponents.
In particular, the diffusion coefficient vanishes with a power law upon approaching the critical density,
\begin{equation}
D\sim |n^*-n^*_c|^\mu.
\label{eq:D_scaling}
\end{equation}
The localization transition was predicted by a mode-coupling approach~\cite{Goetze:1981a,Goetze:1981b,Masters:1982} and by a mapping to continuum percolation~\cite{Kertesz:1983,Machta:1985}.
The direct link between the Lorentz model and continuum percolation,
however, was established only recently~\cite{Lorentz_PRL:2006};
it is substantiated further by the data presented in \sec{simulation_results}.
Since the interpretation of the data depends on an understanding of continuum percolation theory and some of its subtleties, the next section summarizes and motivates the central results of the theory.

\section{Continuum Percolation}
\label{sec:percolation}

One particular interesting class of percolating systems is that of continuum percolation~\cite{Kertesz:1981}, which was later also termed ``Swiss cheese'' model~\cite{Halperin:1985}. Isolating discs or spheres are distributed at random within a conducting material, they may overlap, and the remaining material between them forms the percolation clusters---with respect to the medium, this model is identical to the Lorentz model. In the simplest variant, the spheres are monodisperse, \ie have equal radii, but binary distributions were considered as well~\cite{vdMarck:1996,Rintoul:2000}. The percolation threshold was first determined by discretizing the void space and using lattice methods~\cite{Kertesz:1981}. Much more efficiently, the void space is well represented by a random network constructed from a Voronoi tessellation~\cite{Kerstein:1983}; then, the percolation thresholds are calculated for that network. As expected from the universality hypothesis, the critical exponents for static properties, \eg $\beta$ for the strength of the percolating cluster $P_\infty$ and $\nu$ for the correlation length $\xi$, agree with the findings for various lattices of the same dimension~\cite{Elam:1984}.

The more suprising were theoretical predictions, based on the ``nodes-links-blobs'' picture, that the exponents of transport properties, \eg conductivity and elasticity, are considerably larger than their universal values on lattices~\cite{Halperin:1985,Machta:1985} (with the exception of the conductivity exponent in $d=2$). These results were obtained by assigning physical properties to the bonds of the network, \eg in order to investigate the conductivity, the bonds were considered as resistors. Since the macroscopic conductivity only depends on the probability distribution of the bond conductances, the latter may be distributed randomly obeying a given distribution. A renormalization group analysis of such random resistor networks~\cite{Straley:1982,Harris:1984,Lubensky:1986,Stenull:2001} attributes the non-universal character of the critical exponents to a singularity in the distribution of conductances.
In the following, we will sketch the derivation of such a singular distribution for the Lorentz model, and we will summarize the line of argument for the modification of the tracer dynamics.

\subsection{Mapping to random resistor networks}

\begin{figure*} \hfill
\fbox{\includegraphics[width=.25\textwidth]{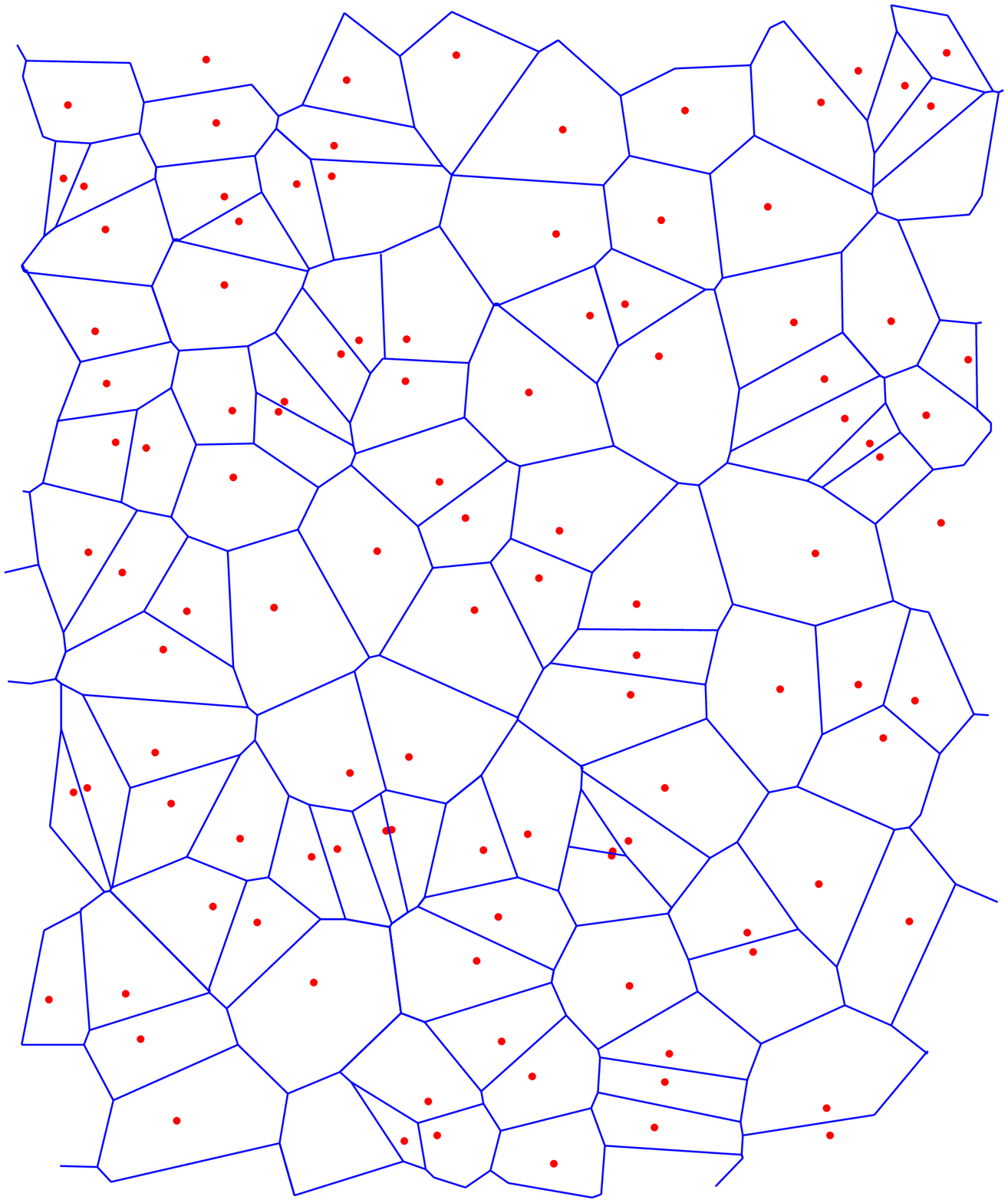}} \hfill
\fbox{\includegraphics[width=.25\textwidth]{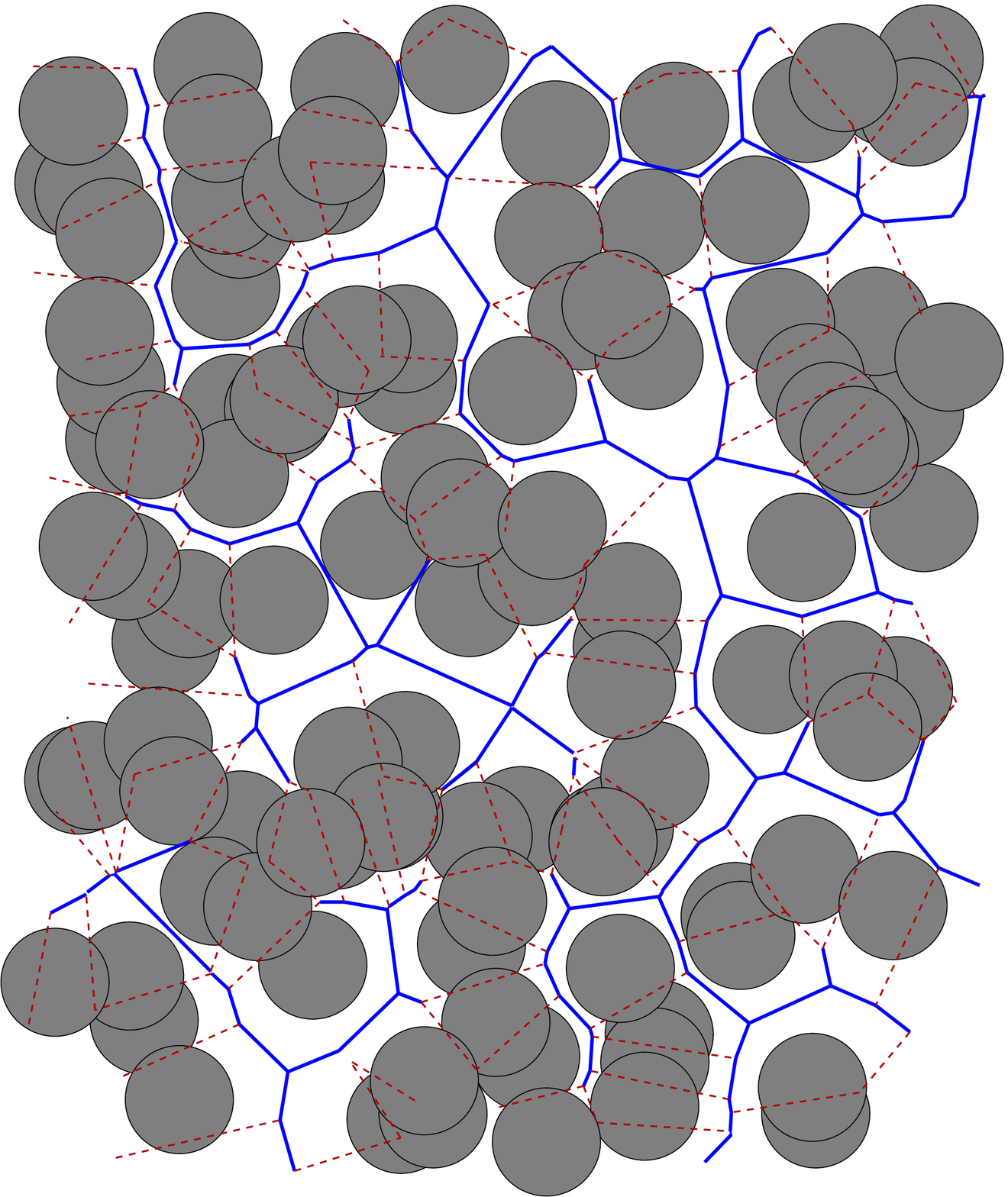}} \hfill
\fbox{\includegraphics[width=.25\textwidth]{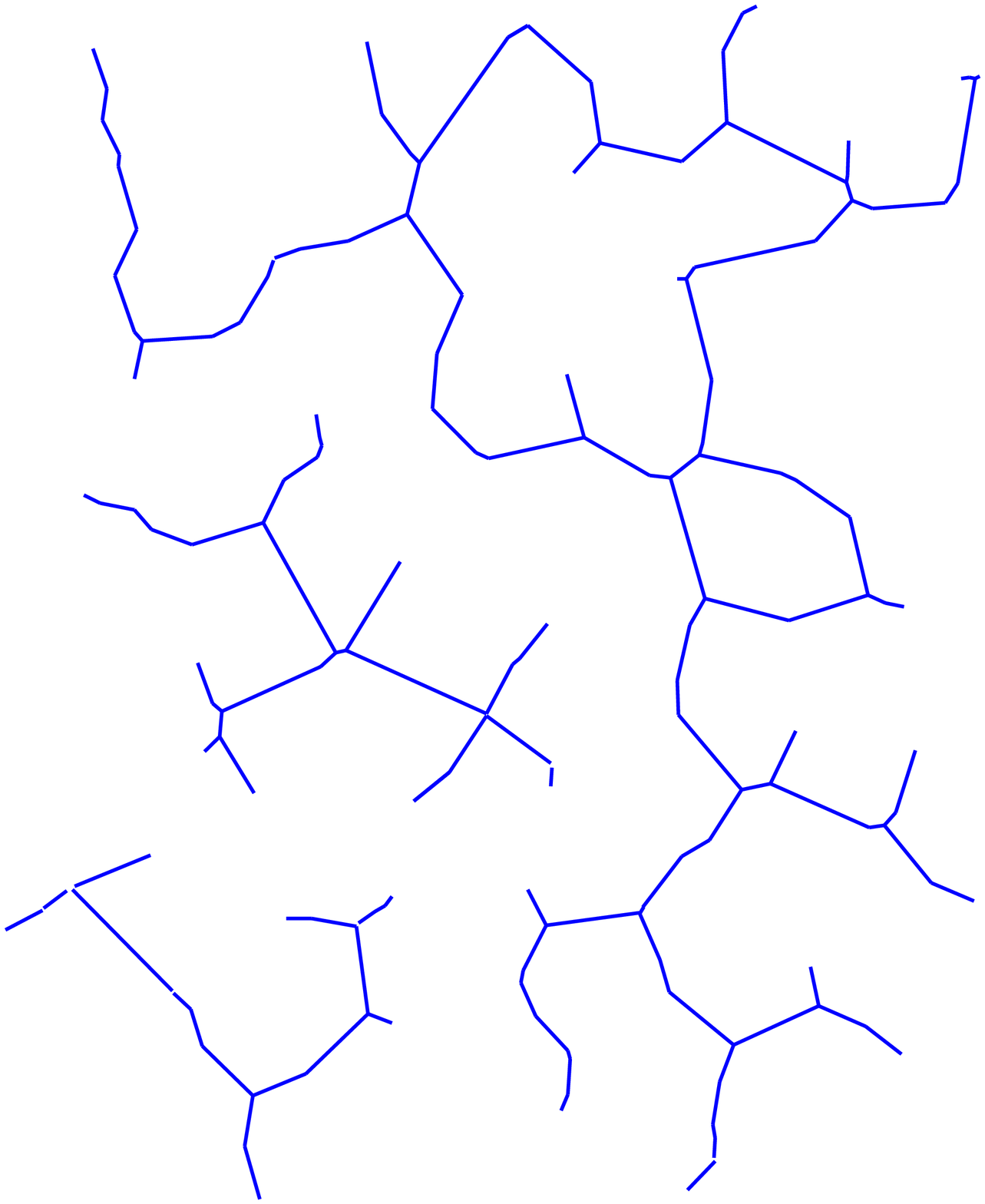}} \hfill\hbox{}
\caption{(Color online) Mapping of the Lorentz model to a random resistor network: (i)~construct the Voronoi tessellation around the obstacle centers, (ii)~remove blocked bonds and assign a conductivity to each bond according to the gap width, and (iii) finally drop the obstacles. (Cartoons are in two dimensions.)}
\label{fig:mapping}
\end{figure*}

The mapping from the Swiss cheese model to a geometric Voronoi network to a random resistor network is indeed intuitive, but not at all rigorous. Further, the ballistic dynamics of the tracer particle in the Lorentz model is quite different from the diffusive dynamics of a random walker on a percolation lattice. Only by virtue of the universality hypothesis, one can expect that both models share the same critical behavior.

For the dynamic properties of the three-dimensional Lorentz model, it is essential that narrow gaps between obstacles are abundant, \ie the gap distribution exhibits a singularity. The mapping from the Lorentz model to a random resistor network includes the following steps (see \fig{mapping}):
\begin{enumerate} \renewcommand{\labelenumi}{(\roman{enumi})}
\item Construct a Voronoi tessellation around the obstacle centers, defining vertices that are connected by edges. Each vertex either defines a ``chamber'' between the obstacles or is outside the void space. Each edge or bond defines a ``gap'' between three obstacles at a time.
(The plane through the obstacle centers separates adjacent chambers.) If a gap is blocked by these obstacles, remove the related bond. Then, any path in the void space can be transformed continuously into a path along the bonds without crossing the obstacles~\cite{Kerstein:1983}.

\item Choose a chamber and one of its gaps. A transition rate $W$ can be assigned to this gap, given as the ratio of the phase-space volume available for leaving through the gap to the total volume $V$ of the chamber. The numerator is proportional to the cross-section area of the gap $A$, and $W\propto A/V$~\cite{Machta:1985}. Slow transport is connected with the distribution $\rho(W)$ for small transition rates. Hence, the singular behavior of $\rho(W)$ for $W\to 0$ is determined by the distribution of small cross-sections $A$ or narrow gaps. In $d$ dimensions, one calculates~\cite{Machta:1985}
\begin{equation}
\rho(W) \sim W^{-\alpha}, \quad \text{where} \quad \alpha=\frac{d-2}{d-1}.
\label{eq:alpha}
\end{equation}

\item Drop the obstacles, keep only vertices and bonds in the void space, and interpret the transition rates through the gaps as conductances along the bonds. This defines the random resistor network with a power-law distribution of weak conductances.
\end{enumerate}
Such random resistor networks have been investigated extensively by means of Monte Carlo simulations~\cite{Derrida:1984,Gingold:1990} and renormalization group techniques~\cite{Harris:1984,Lubensky:1986}, providing reliable numeric and analytic results for the critical behavior~\cite{benAvraham:DiffusionInFractals}.

\subsection{Conductivity}
\label{sec:Conductivity}

The central property of random resistor networks is the (macroscopic) conductivity $\Sigma$. As the infinite cluster becomes increasingly thinner upon approaching the threshold, it vanishes with a power law, $\Sigma\sim|\epsilon|^\mu$, anticipating the same exponent $\mu$ as for the diffusion coefficient due to the celebrated Einstein relation $D\sim \Sigma$~\cite{benAvraham:DiffusionInFractals}.

\begin{figure}
\includegraphics[width=\figwidth]{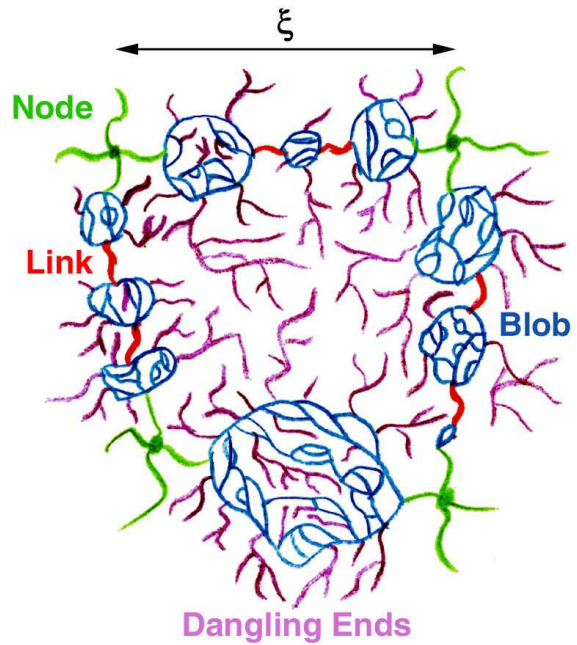}
\caption{(Color online) Cartoon of the nodes-links-blobs model: nodes, being distributed homogeneously with average spacing $\xi$, are connected by chains. A chain is made of a series of links and blobs; links are defined as bonds carrying the whole current of a chain. Most bonds of the network are dangling sites, \ie they carry zero current.}
\label{fig:nodes-links-blobs}
\end{figure}

The percolating network looks homogeneous at length scales larger than a correlation length $\xi$. This fact is pictorially reflected in the ``links, nodes, and blobs'' model by \Citet{Skal:1975} and \Citet{deGennes:1976}; for reviews see~\cite{Nakayama:1994,Stauffer:Percolation}. It simplifies the infinite cluster of the percolating network in the following way (\fig{nodes-links-blobs}): the nodes are defined as vertices which can not be isolated from the cluster by cutting any two (not necessarily adjacent) bonds. The ``nodes'' are supposed to be distributed homogeneously with an average spacing of the correlation length $\xi$. Most vertices of the network belong to cul-de-sacs (``dangling ends''), \ie they can be separated by cutting a single bond somewhere in the network. Nodes are connected by ``chains'', \ie objects that can be isolated by cutting two bonds adjacent to a node. A fraction of the bonds of a chain carries the whole current (``links'' or ``red bonds''), the remaining bonds of a chain form ``blobs'' which connect two links at a time.

The chains have an average resistance $\mathcal{R}$, which exhibits a power-law divergence with exponent $\zeta$ upon approaching the percolation threshold,
\begin{equation}
\mathcal{R}\sim|\epsilon|^{-\zeta}.
\end{equation}
Here we have introduced the separation parameter $\epsilon$; in case of the Lorentz model, we use $\epsilon:=(n^*-n^*_c)/n^*_c$.
Applying an electric field $E$, the potential drop between two nodes is of order $\xi E$ yielding a chain current $I=\xi E/\mathcal{R}$. Then, the current density in the network is given by $j=I/\xi^{d-1}$. Plugging in the definition of the conductivity, $j=\Sigma E$, yields $\Sigma\sim \xi^{2-d}/\mathcal{R}$, and a hyperscaling relation follows,
\begin{equation}
\mu=(d-2)\nu+\zeta.
\label{eq:mu_zeta_hyperscaling}
\end{equation}

The remaining task is to identify the exponent $\zeta$ for such random resistor networks that are relevant for the Lorentz model, \ie that exhibit a power-law distribution of the conductances, $\rho(W)\sim W^{-\alpha}$ with $0\le \alpha < 1$. We will follow the argumentation of \Citet{Straley:1982}. The distribution of chain conductances $\rho_\text{chain}(W)$ can be considered as the renormalized distribution of bond conductances $\rho(W)$ in the sense of the real space renormalization group. The renormalization flow shifts the whole distribution; usually, the peak of the renormalized distribution $\rho_\text{chain}(W)$ is determined by the peak of the microscopic distribution $\rho(W)$. The low-conductivity tail, however, is shifted as well. Two different scenarios arise dependent on the value of $\alpha$: either the peak absorbs the tail, yielding the universal form of the renormalized distribution,
and the chain resistance exhibits the universal exponent $\zeta^\text{univ}$.
In the second case, the tail pulls weight out of the peak, leading to a final distribution with the same small-$W$ tail, $\rho_\text{chain}(W)\sim W^{-\alpha}$. Then, the chain resistance is dominated by the small conductances.
The total resistance of parallel bonds, \ie the blobs, may be neglected compared to the resistance of the red bonds, and $\rho_\text{chain}(W)$ obeys for small $W$
\begin{equation}
\rho_\text{chain}(W)\sim N_\text{red}\rho(W),
\label{eq:rho_chain}
\end{equation}
where $N_\text{red}$ is the number of red bonds within a chain. Dimensional analysis suggests the scaling form
\begin{equation}
\rho_\text{chain}(W)=W_0^{-1}\tilde\rho_\text{chain}(W/W_0),
\end{equation}
with the typical chain conductance $W_0$. From \eq{rho_chain} one infers $N_\text{red}\sim W_0^{\alpha-1}$, and the average chain resistance $\mathcal{R}=\expect{1/W}^{-1}$ is identified as $W_0$. For the divergence of the number of red bonds, we finally employ a result by \Citet{Coniglio:1981},
\begin{equation}
N_\text{red}\sim|\epsilon|^{-1}.
\end{equation}
Collecting results,
\begin{equation}
\mathcal{R}\sim W_0 \sim N_\text{red}^{1/(\alpha-1)} \sim |\epsilon|^{1/(1-\alpha)},
\end{equation}
yields  $\zeta = (1-\alpha)^{-1}$. It has been argued that the crossover between both scenarios occurs such that~\cite{Straley:1982,Machta:1986}
\begin{equation}
\zeta = \max\left[(1-\alpha)^{-1},\zeta^\text{univ}\right].
\end{equation}

Without resorting to the assumptions of the ``links, nodes, and blobs'' model, \Citet{Stenull:2001} have proven directly by means of an expansion of the renormalization group equations in $\epsilon=6-d$ to arbitrary order the equivalent relation
\begin{equation}
\mu = \max\left[(d-2)\nu+(1-\alpha)^{-1},\mu^\text{lat}\right],
\label{eq:mu_hyperscaling}
\end{equation}
where $\mu^\text{lat}$ is the universal exponent for lattice percolation.

From simulations, it is known that $1\le \zeta^\text{univ}\lesssim 1.3$ for \mbox{$d\ge 2$} dimensions~\cite{Stauffer:Percolation}. Using the above value of $\alpha$ for the Lorentz model, \eq{alpha}, it follows that $\zeta$ takes its universal value on lattices only for $d=2$ dimensions, otherwise $\zeta=d-1$.

\subsection{Transport on percolation networks}

Going back to the Lorentz model in the series of mappings, the transport of a particle in the percolating void space may be understood as a random hopping process between the nodes of the network with lattice constant $\xi$. Hence, the motion of such a particle is diffusive at scales much larger than the correlation length $\xi$. Note that $\xi$ also characterizes the linear dimension of the largest finite clusters. Along the fractal chains of the network, however, the dynamics is slow and exhibits anomalous transport characterized by the walk dimension $\dw>2$,
\begin{equation}
  \expect{\Delta \vec R(t)^2}_\text{chain}\sim t^{2/\dw}.
\end{equation}
The time scale associated with $\xi$ obeys $\tX \sim \xi^\dw$, and diffusion is observed for long times, $t\gg \tX$, with a coefficient according to \eq{D_random_walk}. If the particle starts at \emph{any} cluster, the diffusion coefficient is given by
\begin{equation}
  D = P_\infty\, \xi^2/\tX \sim |\epsilon|^\beta \xi^{2-\dw},
  \label{eq:D_xi_scaling}
\end{equation}
where the factor $P_\infty$ allows for the zero diffusion coefficient of particles on finite clusters.
Recalling $D\sim|\epsilon|^\mu$ and $\xi\sim|\epsilon|^{-\nu}$, the conductivity exponent
is related to the walk dimension by
\begin{equation}
  \mu = (\dw-2)\nu + \beta.
\end{equation}

In the above calculation, the diffusion coefficients of particles from different clusters were averaged. Instead of taking the cluster average in the end, one may average already the mean-square displacement.
Then, the relevant dynamic length scale is the root-mean-square cluster radius~\cite{Lorentz_PRL:2006,PhD_thesis:2006},
\begin{equation}
  \label{eq:ell_scaling}
  \ell\sim|\epsilon|^{-\nu+\beta/2},
\end{equation}
and anomalous transport is characterized by the dynamic exponent $z$
different from $\dw$~\cite{benAvraham:DiffusionInFractals},
\begin{equation}\label{eq:msd_critical}
  \delta r^2(t)\sim t^{2/z} \quad \text{for} \quad t\ll \tX.
\end{equation}
The crossover time scale $\tX$, being the same for all clusters, is not affected by the cluster average. It holds $\ell^z\sim\xi^\dw\sim \tX$,
and therewith $D\sim\ell^2/\tX\sim\ell^{2-z}$, which yields a scaling relation connecting $\mu$ and $z$,
\begin{equation}
  z=\frac{2\nu-\beta+\mu}{\nu-\beta/2}.
  \label{eq:z_scaling}
\end{equation}
In particular, this relation implies a connection of the exponents $z$ and $\zeta$ via \eq{mu_zeta_hyperscaling}, and $z$ is expected to deviate from its universal value on lattices for $d\ge 3$. Moreover, $z$ can be calculated from the geometric exponents $\nu$ and $\beta$ which are believed to equal their universal lattice values~\cite{Elam:1984}; in three dimensions, we use $\nu=0.88$ and $\beta=0.41$~\cite{Stauffer:Percolation} throughout this work, evaluating \eq{z_scaling} to $z=6.25$.

\section{Simulation details}
\label{sec:simulations}

Molecular Dynamics simulations allow for a direct numerical analysis of the dynamic properties of the Lorentz model without resorting to random resistor networks. Thus, a quantitative description over the full density range becomes accessible. Ballistic trajectories are produced by means of a standard simulation algorithm already employed by \Citet{Bruin:1972}. It is combined with a method for calculating correlation functions online, optimized for exponentially large time scales. For Brownian particles, we have extended this simulation algorithm to include stochastic forces similar to recently discussed ideas~\cite{Scala:2007}.

\subsection{Ballistic particles}
\label{sec:ball-lorentz-model}

We use an event-oriented simulation algorithm, since the tracer particle and the obstacles interact via a hard-core potential. The algorithm propagates the particle freely from collision to collision, in each case calculating the precise point in time of the next interaction with an obstacle. If the particle is located at $\vec r$ with velocity $\vec v$, it possibly hits a single obstacle in the coordinate origin after the time interval~\footnote{The numerical error can be reduced using $t_\text{coll}=\min[q/v^2,(r^2-\sigma^2)/q],$ where $q=-b+\sqrt{b^2-v^2(r^2-\sigma^2)}$ and $b<0$. This formula avoids the calculation of the difference between two almost equal numbers which would occur for $b^2\gg v^2(r^2-\sigma^2)$.}
\begin{equation}
t_\text{coll} = -\frac{b}{v^2} - \frac{1}{v^2}\sqrt{b^2-v^2(r^2-\sigma^2)},
\label{eq:collision_time}
\end{equation}
where $b=\vec v\dotprod \vec r$.  If $b>0$, the
particle departs from the obstacle and no collision will take place.
If the radicand becomes negative, the particle misses the obstacle.
The particle is scattered specularly, its post-collisional velocity is
\begin{equation}
\vec v'=\vec v - 2(\vec v \dotprod \unitvec \sigma)\, \unitvec \sigma,
\label{eq:law_of_collision}
\end{equation}
where $\unitvec \sigma = (\vec v t_\text{coll} +\vec r)/\sigma$
specifies the surface normal at the collision point.

Therewith, the implemented algorithm is straightforward: In a preparation step, the obstacle
positions and the initial phase space coordinates of the particle
are chosen randomly from a uniform distribution, under the constraint that the
particle starts in the void space and $|\vec v|=v$.
To reduce the number of collision tests, the simulation box is divided into small cubic cells,
on average containing one or two obstacles; only obstacles from the
cell where the particle is located and from neighboring cells are considered.
Then, repeated collision tests and propagation of the particle to the
next collision point yield the trajectory.

Since a single trajectory can consist of several billion collisions, one easily
runs out of computer memory if one would store the complete trajectory. An
efficient blocking scheme \linebreak (``order-n algorithm'', see~\cite{Frenkel:MD})
takes care of this issue by arranging the trajectory on a
logarithmic time grid and simultaneously calculating various correlation
functions $C(t;t_0)$. The algorithm already averages over different time origins
$t_0$ (``moving time average''), which are, however, not necessarily
uncorrelated, especially for short time intervals $t-t_0$. Hence, we will not
infer any estimate of the statistical error from this averaging procedure.
Rather, a set of $N_t$ trajectories with different initial positions for each of
$N_r$ different realizations of the obstacle disorder is simulated, and the
statistical error is estimated from $N_t\times N_r$ independent measurements.
 At each density, we have simulated at least $N_r=20$ realizations of the disorder.
At intermediate densities, the total number of trajectories has been chosen
$N_t\times N_r>100$. This value has been increased up to 600 at very high
densities, where the phase space is highly decomposed into small, disconnected
parts.
The longest trajectories span about 10\textsuperscript{10} collisions, the demand on
CPU time for such a trajectory was about 15~hours on a single AMD Opteron 248 processor core.

All numerical results presented in this work refer to fixed dimensionality,
$d=3$. The simulation box has periodic boundaries, and its linear size was chosen as
$L=200\sigma$. A detailed finite-size analysis is presented in
\sec{finite_size}.

\subsection{Brownian particles}
\label{sec:brownian-dynamics}

When the pores between the obstacles are filled with some solvent, the tracer
particle performs Brownian motion on microscopic time and length scales. 
By another Einstein relation, the solvent's friction coefficient and temperature
yield the coefficient $D_0^\text{B}$ of free diffusion.
It may be combined with an intrinsic length scale in the system, \eg the obstacle radius~$\sigma$, yielding a microscopic time scale $t_0:=\sigma^2/D_0^\text{B}$.
All averages for Brownian particles will refer to the canonical ensemble throughout the article.

\begin{figure}
  \centering
    \includegraphics[angle=90,width=.85\figwidth]{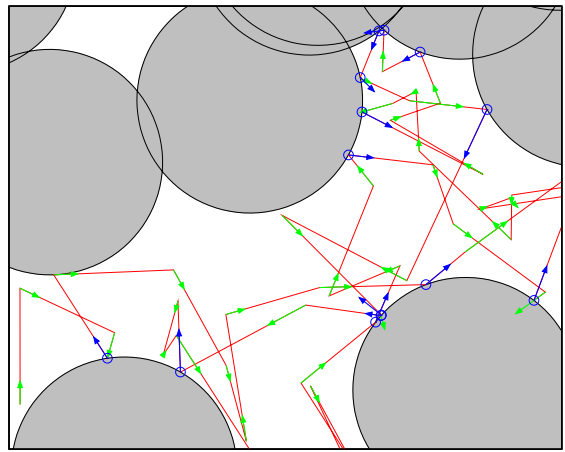}
    \caption{(Color online) Typical particle trajectory in two dimensions, demonstrating the
      Brownian dynamics algorithm. Collisions with an obstacle are indicated by
      blue circles. Arrows indicate the direction and magnitude of the velocity
      after a collision (blue/dark) and after drawing a new velocity from the
      Boltzmann distribution (green/gray).}
  \label{fig:brown_t+b}
\end{figure}

The overdamped dynamics of the tracer may be described by an effective
stochastic force, caused by incessant collisions with solvent molecules.
We have extended the above simulation algorithm by taking advantage of a
coarse-grained scheme, where the ballistic trajectory is frequently interrupted
by an equilibration with the solvent.
After each fixed time interval $\tauB$ a new velocity is assigned to the tracer
particle, randomly drawn from the normalized Boltzmann distribution, 
\begin{equation}
  \label{eq:Boltzmann}
  p_{\vec v}(\vec v)\propto \exp\left(-\frac{mv^2}{2\kB T}\right).
\end{equation}
Any hydrodynamic interactions are neglected.
Such a simulation scheme for Brownian particles with hard-core interactions was 
carefully
tested recently~\cite{Scala:2007}; a similar approach was already studied
by~\textcite{Alley:1979}, where the equilibration was restricted to the instants
of particle collisions. \fig{brown_t+b} exemplifies the motion in two
dimensions, demonstrating the velocity changes along the trajectory caused by
events of collision and equilibration.

\begin{figure}
  \includegraphics[width=\figwidth]{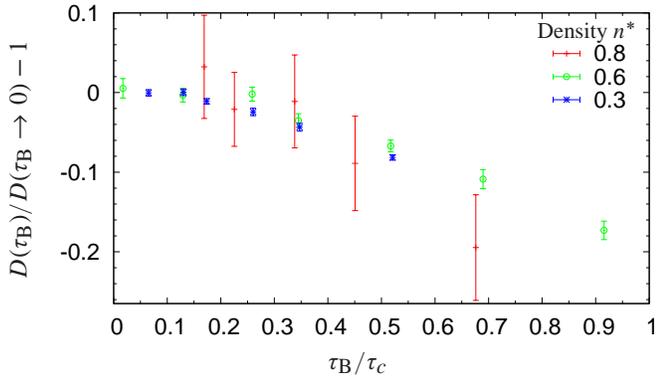}
  \caption{(Color online) Convergence of the macroscopic diffusion coefficient of the Brownian
    particle for decreasing time interval $\tauB$ at different densities. The
    limits $D(\tauB\to0)$ are estimated by averaging over results for
    $\tauB\leq 0.15\,v^{-1}\sigma$.}
  \label{fig:diffconst_B}
\end{figure}

Without obstacles, the algorithm yields a $\tauB$-dependent diffusion coefficient, cf.~\eq{D_random_walk},
\begin{equation}
  \label{eq:D0_brownian}
  D_0^\text{B} = \frac{\expect{v^2}\tauB}{2d} = \frac{\kB T}{2m} \tauB.
\end{equation}
The obtained motion is obviously only diffusive for times larger than $\tauB$.
To model Brownian motion at all physically relevant time scales, one has to
choose $\tauB$ smaller than the shortest time scale of the system;
in the present problem, this is the average time interval $\tau_c$ between collisions with obstacles.
The dependence of the diffusion coefficient on $\tauB$ is shown in
\fig{diffconst_B}. We consider the value of $\tauB$ sufficiently small when the macroscopic
diffusion coefficient does not depend on $\tauB$ anymore, resulting in $\tauB\approx0.3\tau_c$ at criticality, $n^*=0.839$.

\begin{figure}
\includegraphics[width=\figwidth]{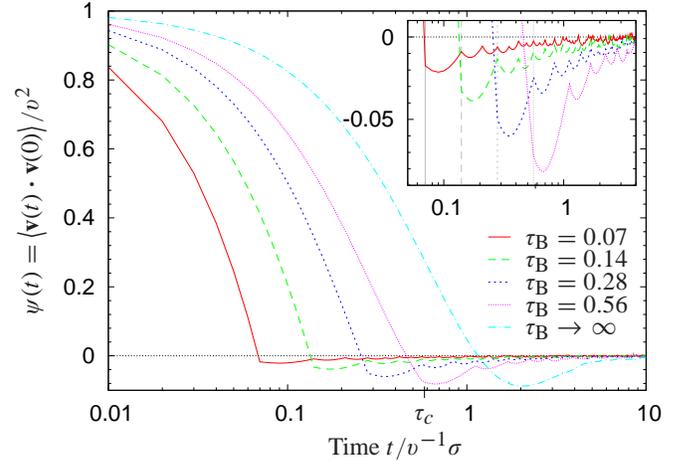}
\caption[Velocity-autocorrelation function]{(Color online) Velocity-autocorrelation function $\psi(t)$ obtained from the
  Brownian dynamics algorithm, at density $n^*=0.6$. Simulation results are
  shown for four different time intervals $\tauB$ and for the limiting case
  $\tauB\to\infty$, equivalent to a ballistic particle. $\psi(t)$ does not vanish
  immediately for $t>\tauB$, but oscillates with frequency $1/\tauB$. On the
  $x$-axis, the mean collision rate $\tau_c^{-1}$ is also indicated, varying
  inappreciably with $\tauB$.
  Inset: magnification of the velocity anti-correlations; vertical gray lines
  indicate $\tauB$ of the data with the corresponding dashing pattern.
  Note that the chosen $\tauB$ are multiples of the smallest one.}
\label{fig:vacf-d06}
\end{figure}

Considering the velocity autocorrelation function $\psi(t)=\expect{\vec
  v(0)\dotprod\vec v(t)}/v^{2}$ for the combined algorithm of Brownian dynamics and
ballistic collisions, one would naively expect that all correlations in
the velocity vanish immediately for times $t>\tauB$. This seems reasonable,
since a new velocity drawn at random should be
uncorrelated to the previous value. The presence of excluded volume, however,
induces correlations for particles sufficiently close to an obstacle if the
Brownian update interval and the mean interval between collisions are comparable,
$\tauB\approx\tau_c$. As a consequence, $\psi(t)$ shows periodic
anticorrelations for times $t>\tauB$, with the periodicity interval $\tauB$
and an exponentially decaying amplitude. This effect is demonstrated in \fig{vacf-d06}
for several different values of $\tauB$. 
Since the focus of this work is on the long-time properties, we defer a more detailed explanation to Appendix~\ref{sec:short-time-velocity}.

\section{Simulation results}
\label{sec:simulation_results}

\begin{figure*}
  \includegraphics[width=\textwidth]{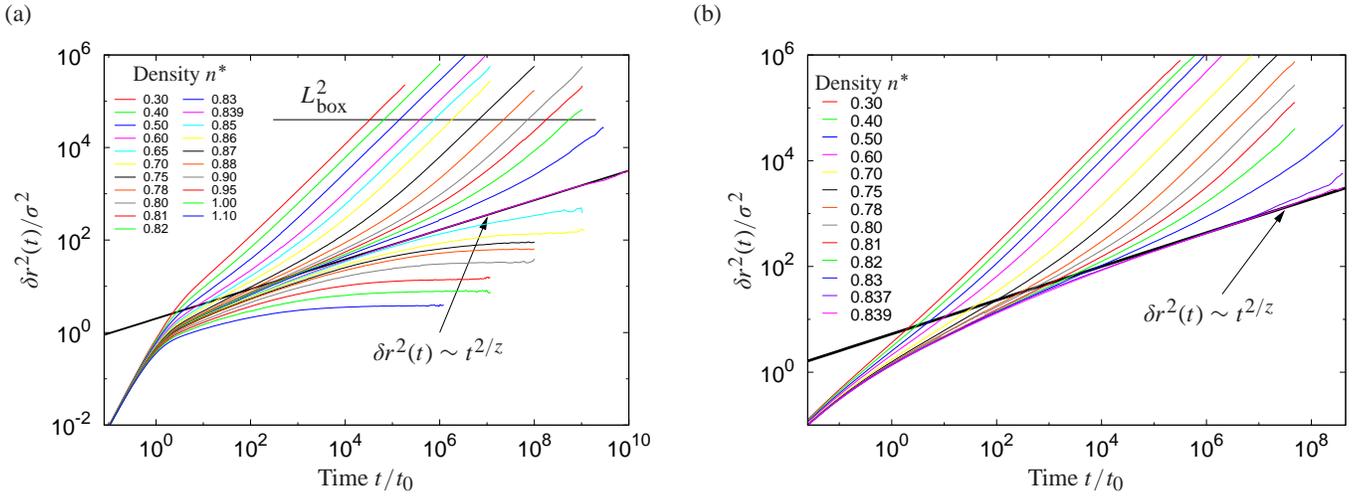}
  \caption{(Color online) Mean-square displacement $\delta r^2(t)$ of the Lorentz model with
    (a) ballistic and (b) Brownian particles. The obstacle density $n^*$
    increases from top to bottom; thick black lines indicate the long-time asymptote
    for anomalous transport at criticality, $\delta r^2(t)\sim t^{2/z}$ with $z = 6.25$.}
  \label{fig:msd3d}
\end{figure*}

We have simulated trajectories of ballistic and Brownian tracer particles over a wide range of
obstacle densities, above and below the localization transition (\fig{msd3d}).
The mean-square displacement exhibits three distinct regimes: at short times,
$t\ll t_0$, transport is not hindered by the obstacles, and the intrinsic dynamics
of the particle is observed, either ballistic or Brownian.
In both cases, this regime is followed by a regime of anomalous transport to be discussed in detail
later. At large time scales, $t\gg \tX$, and below the localization transition,
 generic diffusive behavior is recovered.
Above the critical density, the particles are trapped and $\delta r^{2}(t)$
is bounded by $\ell^2$, the mean-square cluster size. At the critical density,
the dynamics becomes neither diffusive nor trapped, and transport remains anomalous for all times.

\subsection{Ballistic particles}

\begin{figure*}
\includegraphics[width=\textwidth]{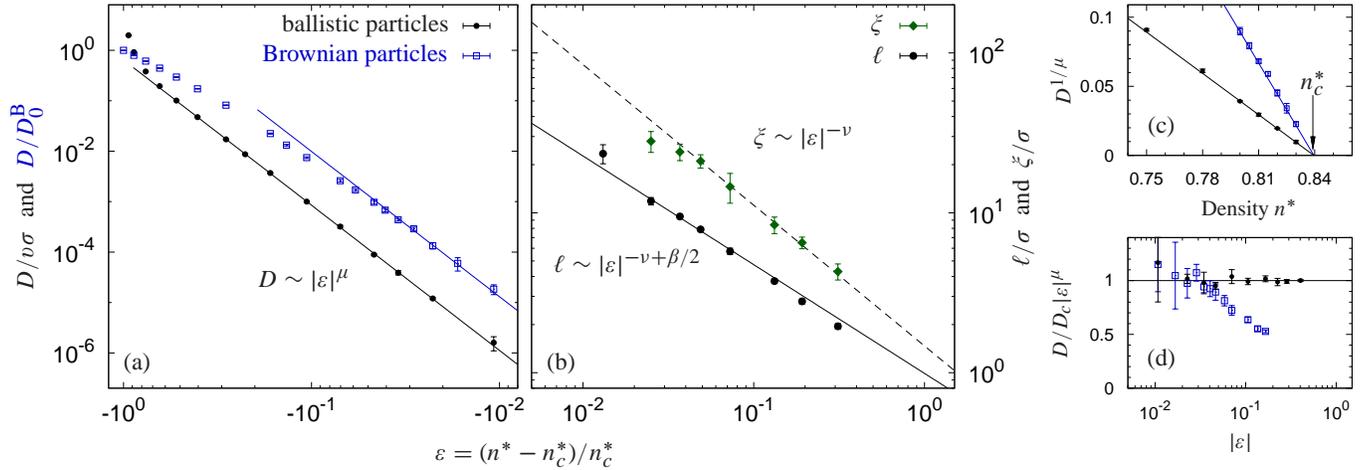}
\caption[Power laws of $D$, $\ell$ and $\xi$ close to the transition]{
  (Color online)
  Critical power law behavior close to the localization transition;
  black closed symbols refer to ballistic particles, blue open squares to Brownian ones.
  ~~(a)~As the
  localization transition is approached, the diffusion coefficient $D$ vanishes
  with exponent $\mu=2.88$. Units are $v\sigma$ for ballistic and $D_0^\text{B}$ for Brownian particles.
  ~~(b)~The localization length $\ell$ diverges with
  exponent $\nu-\beta/2$ and can clearly be distinguished from the correlation
  length, $\xi\sim|\epsilon|^{-\nu}$.
  ~~(c)~Rectification of the diffusion coefficient close to the critical density. Fitting a straight line to the closest five data points yields the critical density $n^*_c$.
  ~~(d)~Comparing $D$ with its asymptotic behavior $D_c |\epsilon|^\mu$ on a semi-logarithmic scale reveals that the deviation from the power law is less than $10\%$ for ballistic particles, except for the smallest data point.}
\label{fig:D_ell_xi3d}
\end{figure*}

Let us discuss ballistic particles first.
Diffusion coefficients were extracted from the mean-square displacements,
$D=\partial_t \delta r^2(t\to\infty)/6$. They vanish as a critical density $n^*_c$ is approached from
below according to a power law, $D\sim|n^*-n^*_c|^\mu$.
The exponent $\mu$ could be fitted to
these data; such a fit, however, depends on a simultaneous determination of the
critical density~$n^*_c$. We will rather use a theoretical prediction for
$\mu$ that relies on the mapping to random resistor networks; therewith, a first
test of the validity of the mapping is obtained. \eq{mu_hyperscaling} together
with the result $\alpha=1/2$ from \Citet{Machta:1985}, \eq{alpha}, provides
$\mu=\nu+2\approx 2.88$. This prediction is clearly corroborated by our results:
in \fig{D_ell_xi3d}, the diffusion coefficient obeys the power law over five orders of magnitude, with a deviation of less than 15\%.

The knowledge of $\mu$ allows to fit the critical density quite precisely
by means of a rectification plot, showing
$D^{1/\mu}$ against $n^*$ on a linear scale; see \fig{D_ell_xi3d}c.
We obtain $n^*_c=0.839(4)$, which coincides with the percolation threshold of
the void space, $n^*_\text{perc}=0.8363(24)$~\cite{Kertesz:1981,Elam:1984,vdMarck:1996,Rintoul:2000}.
The interpretation is
that the particle will eventually squeeze through any gap, no matter how narrow.
There are no regions
on a cluster which are too improbable to be visited by the particle after an
infinitely long time. This means furthermore that the particle will diffuse as
long as its surrounding phase space is connected with infinity.

In the localized regime, $n^*>n^*_c$, the long-time limit of the mean-square
displacement directly yields the mean cluster radius, $\delta
r^2(t)\simeq \ell^2$ for $t\gg \tX$. Then, the correlation length~$\xi$ is
easily accessible upon observing that the
mean quartic displacement $\delta r^4(t):=\expect{\Delta \vec
  R(t)^4}$ scales for long times as~\cite{Lorentz_PRL:2006}
\begin{equation}
  \delta r^4(t) \simeq \xi^2 \ell^2 \quad \text{for} \quad t\gg \tX;
  \label{eq:mqd_longtimelimit}
\end{equation}
this relation is taken as definition for $\xi$.
The extracted length scales $\ell$ and $\xi$ diverge at the localization transition
with exponents $\nu-\beta/2=0.68$ and $\nu=0.88$, respectively, according to Eqs.~\eqref{eq:ell_scaling} and \eqref{eq:D_xi_scaling}; see \fig{D_ell_xi3d}b. The values for the exponents are taken from lattice percolation~\cite{Stauffer:Percolation}, and thus our data corroborate that the geometric exponents from lattices apply equally to continuum percolation~\cite{Elam:1984}.
Further, the localization length $\ell$ is identified with the mean-cluster size, as already anticipated by our notation, and clearly contrasted with the correlation length $\xi$.
Both length scales are relevant for the dynamics as proposed in Ref.~\cite{Lorentz_PRL:2006}.
In particular, our results corroborate the interpretation of the dynamic properties of the Lorentz model in terms of
random walks on percolation clusters.

\begin{figure}
  \includegraphics[width=\figwidth]{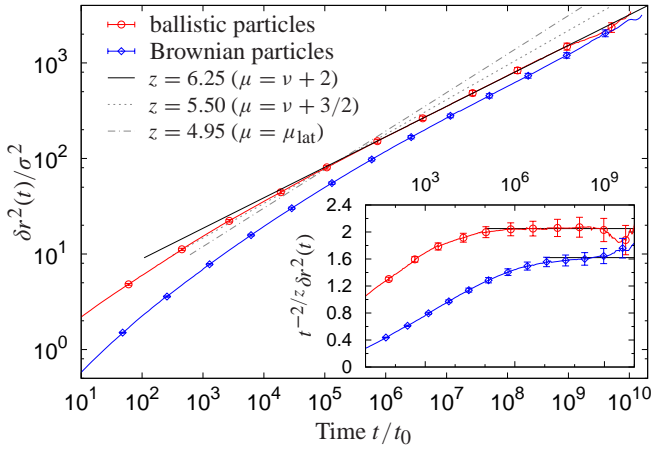}
  \caption[Critical behavior of the mean-square displacement]{(Color online) Critical behavior
    of the mean-square displacement at $n^*=0.839\approx n_{c}$ for ballistic and
    Brownian particles. Straight lines represent power laws $t^{2/z}$, where $z$
    is related to $\mu$ by \eq{z_scaling}. Three predicted values of $\mu$ are
    compared: the scaling relations $\mu=\nu+2$~\cite{Machta:1985} and
    $\mu=\nu+3/2$~\cite{Halperin:1985} from \eq{mu_hyperscaling} as well as the
    value on lattices, $\mu^\text{lat}=2.0$. Our data are compatible only with
    the prediction by \Citet{Machta:1985}. Inset: as a most sensitive test, the
    same data are rectified employing $\mu=\nu+2$; straight
    lines indicate the estimated long-time limits. The convergence for long
    times provides compelling support for this value of $\mu$.}
  \label{fig:msd-crit+rect}
\end{figure}

There is a competing prediction $\mu=\nu+3/2$~\cite{Halperin:1985}, which is also found in text books~\cite{benAvraham:DiffusionInFractals}. A third, maybe naive guess suggests
$\mu^\text{lat}\approx 2.0$, the universal value for lattice percolation in $d=3$~\cite{Stauffer:Percolation}.
We will allow for these alternative predictions by
a sensitive and unbiased test based on \eq{z_scaling},
which relates the different values of $\mu$ to different exponents
$z$ for the anomalous transport. The latter can be inferred directly at the critical density $n^*=0.839$
from the subdiffusive behavior of the mean-square displacement, growing as $\delta r^2(t)\sim t^{2/z}$ for long times, see \fig{msd-crit+rect}.
In the double-logarithmic plot, our data exhibit a slope manifestly smaller than expected from the alternative values for $z$. The data, however, collapse very well with a slope $2/z=2/6.25$ corresponding to $\mu=2.88$. For $t>10^5 \sigma/v$, we find a deviation of the mean-square displacement from this asymptotic behavior by less than 7\% over time scales spanning 4~decades, see inset.
In conclusion, only the value $\mu=2.88$ is consistent
with our data, the other two candidates can clearly be ruled out. Hence, the
hyperscaling relation, \eq{mu_hyperscaling}, with the value $\alpha=1/2$ holds
for the Lorentz model.

\subsection{Brownian particles}
\label{sec:brownian_dynamics}

For Brownian particles, the different microscopic dynamics becomes immediately apparent in the short time regime of the mean-square displacement, see \fig{msd3d}b. 
While ballistic particles show a pronounced bending to diffusive or subdiffusive motion at the time scale of the collisions $\tau_c$, Brownian particles are hardly effected by the obstacles at low densities. At higher densities, the excluded volume induces a transient subdiffusive regime again; the crossover from microscopic diffusion to the subdiffusive regime, however, is considerably more spread out.

From universality arguments one expects that the diffusion coefficient vanishes again at some critical density with a power law that has the same exponent $\mu$ as in the case of ballistic particles.
Plotting $D^{1/\mu}$ against density $n^*$ indeed yields a straight line, see \fig{D_ell_xi3d}c. In addition, one infers that the critical density for Brownian and ballistic particles is the same, emphasizing that the transition is due to geometric rather than dynamic effects. Yet the asymptotic power law is markedly slower approached as in the ballistic case.

The slow convergence for overdamped microdynamics is mostly pronounced in the mean-square displacement at criticality, see \fig{msd-crit+rect}. For $t\approx 10^6 t_0$, it is still more than 10\% off---about a factor 100 slower than for ballistic particles. Nevertheless, anomalous diffusion with the same dynamic exponent $z$ is ultimately observed.

\begin{figure}
  \includegraphics[width=\figwidth]{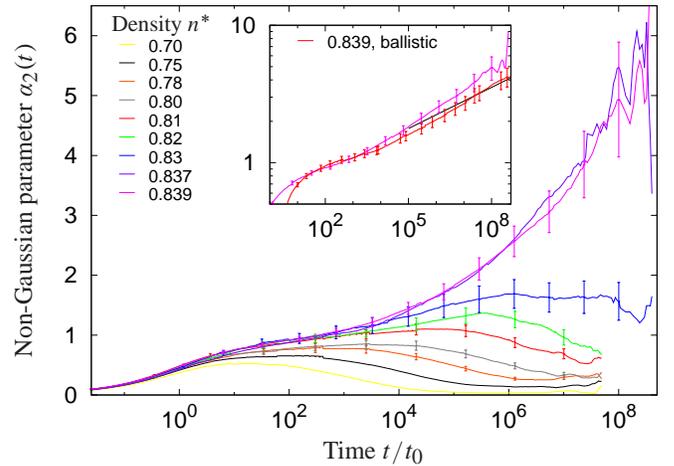}
  \caption{(Color online) The non-Gaussian parameter for Brownian particles yields finite values in the long-time limit, and it increases substantially as the localization transition is approached. Inset: at criticality, $\alpha_2(t)$ grows with a power law in time; the predicted exponent $0.097$ is indicated by a straight line.}
  \label{fig:ngp-brownian_B015}
\end{figure}

The spatially heterogeneous character of the medium renders the motion non-Gaussian, \ie the distribution of particle displacements after a given time lag deviates from a Gaussian distribution. In supercooled liquids, dynamic heterogeneities have been quantified in terms of the non-Gaussian parameter~\cite{Kob:1997}, defined as~\cite{BoonYip:1980}
\begin{equation}
 \alpha_2(t):=\frac{3}{5} \frac{\delta r^4(t)}{[\delta r^2(t)]^2}-1.
\end{equation}
At moderate densities, the presence of the obstacles lets $\alpha_2(t)$ rise to values around $1$ at intermediate times, until it decays to a finite value close to zero, see \fig{ngp-brownian_B015}. This finite long-time limit is due to particles trapped in finite clusters; it diverges as criticality is approached. At criticality, the non-Gaussian parameter is predicted to grow with a power law, $\alpha_2(t)\sim t^{\beta/(2\nu-\beta+\mu)}$, as a consequence of the competition between the localization length $\ell$ and the correlation length $\xi$~\cite{Lorentz_PRL:2006}. Our data provide evidence for a strong increase of $\alpha_2(t)$ close to $n^*_c$, and a double-logarithmic plot yields straight lines. The predicted exponent is very small, $\beta/(2\nu-\beta+\mu) = 0.097$, and difficult to observe, but compatible with our data.

\subsection{Dynamic scaling}

\begin{figure*}
  \includegraphics[width=\textwidth]{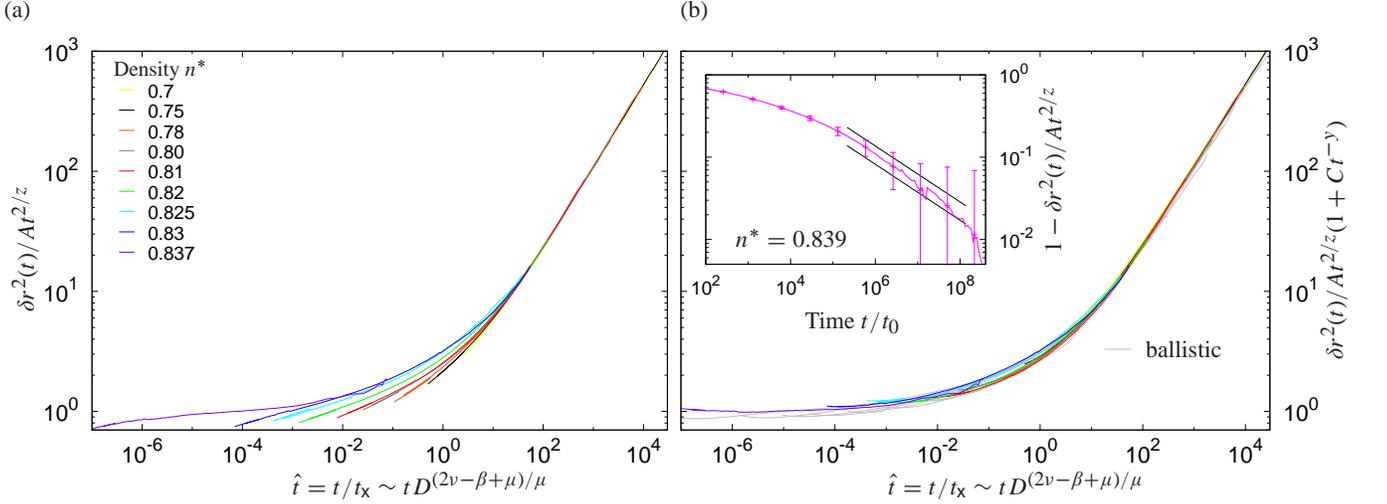}
  \caption[]{(Color online) Scaling plots of the mean-square displacements for Brownian particles. Anomalous transport in the critical regime corresponds to a constant for $\hat t\to 0$. Time is rescaled with the measured diffusion coefficients.
    ~~(a)~Scaling at leading order: as $n^*\to n^*_c$, data converge to 1 for small $\hat t$, and an asymptotic data collapse may be anticipated.
    The amplitude $A$ was determined from the inset of \fig{msd-crit+rect}.%
    ~~(b)~Inclusion of corrections to scaling at leading-order, \eq{msd_scaling_corr} with $\Delta_-(\hat t)= C=-9.5 t_0^y$, yields excellent data collapse onto the scaling function $\widehat{\delta r}{}^2_-(\hat t)$.
    Rescaled and corrected mean-square displacements for ballistic particles ($C=-0.8 t_0^y$) are added in gray from Fig.~2b of Ref.~\cite{Lorentz_PRL:2006}.
    The perfect match of Brownian and ballistic scaling functions substantiates universality of both systems.
    Inset: the corrections at criticality, $n^*=0.839$, decay with a power law for long times, see \eq{msd_critical_corr}. The exponent $y=0.34$ is nicely corroborated over two decades in time, and the amplitude can be limited to the range $-9\leq Ct_0^{-y}\leq -15$ (straight lines).}
  \label{fig:msd-brown-scal}
\end{figure*}

A more stringent test of the universality of ballistic and Brownian particles compares dynamic scaling functions.
The dynamic scaling ansatz for the mean-square displacement reads
\begin{equation}
  \label{eq:msd_scaling}
  \delta r^{2}(t;\epsilon)\simeq t^{2/z}\widehat{\delta r}{}^2_{\pm}(t/\tX)
\end{equation}
for $\epsilon\to 0,\, t\gg \tau_0$, and with the crossover time scaling as $\tX\sim \ell^z$. Note that the relevant length scale for the mean-square displacement is the mean cluster size~$\ell$ rather than the correlation length~$\xi$, due to the infinite life time of the percolation clusters. The subscript $\pm$ at the scaling functions refers to the sign of $\epsilon$, discriminating the different behaviors in the long-time limit. 
In the following, we will restrict the discussion to the diffusive regime, $\epsilon<0$. The scaling function $\widehat{\delta r}{}^2_-(\hat t)$ with $\hat t:=t/\tX$ interpolates between anomalous diffusion at criticality and normal diffusion for long times; thus
\begin{equation}
  \widehat{\delta r}{}^2_-(\hat t)\simeq \begin{cases}
  A & \text{for} \quad \hat t\to 0, \\
  A'\, \hat t^{1-2/z} & \text{for} \quad \hat t\to\infty.
  \end{cases}
\end{equation}
The crossover time may be defined quantitatively by matching both regimes, $A \tX^{2/z}=6 D \tX$, implying that $\tX\sim D^{z/(2-z)}$ for $\epsilon\to 0$. For ballistic particles, the diffusion coefficient observes nicely the asymptotic law $D\sim |\epsilon|^\mu$, see \fig{D_ell_xi3d}d, and plotting $\delta r^2(t)/A t^{2/z}$ vs. $t |\epsilon|^{2\nu-\beta+\mu}$ yields a satisfactory data collapse, see Fig.~2b in Ref.~\cite{Lorentz_PRL:2006}.
For Brownian particles however, our data for the diffusion coefficients are not yet in the asymptotic regime and show significant deviations from this behavior. Rescaling the mean-square displacements with $\tX$ in terms of $\epsilon$ is not expected to lead to data collapse. Instead, let us rescale time with the measured diffusion coefficients, $\tX\propto D^{-(2\nu-\beta+\mu)/\mu}$, allowing for the deviations of $D$ from its asymptotic behavior, see \fig{msd-brown-scal}a; the exponent evaluates to $(2\nu-\beta+\mu)/\mu\approx 1.47$. All curves collapse in the diffusive regime ($\hat t\gg 1$) by construction; but in the critical regime ($\hat t\ll 1$), the data fan out, and asymptotic convergence to a constant may only be anticipated.

Such a behavior hints at corrections to the leading scaling behavior, which carry some fading reminiscence of the microscopic structure. One has to distinguish between analytic corrections, which depend, e.g., on the choice of the separation parameter, and universal corrections with non-integer powers (or logarithmic terms). In the case of ballistic particles~\cite{Lorentz_PRL:2006}, we have identified the leading correction as a universal power law, and excellent data collapse has been achieved in the critical regime too by allowing for these terms. We have argued that the scaling ansatz for the mean-square displacement including the leading correction reads
\begin{equation}
  \label{eq:msd_scaling_corr}
  \delta r^2(t;\epsilon)\simeq t^{2/z}\widehat{\delta r}{}^2_\pm(\hat t)\left[1+t^{-y}\Delta_\pm(\hat t)\right].
\end{equation}
The exponent $y$ is universal and will be discussed below. The correction function $\Delta_\pm(\hat t)$ is universal too, but unknown. At criticality, it reduces to a constant, $\Delta_\pm(0)=:C$, and the mean-square displacement obeys
\begin{equation}
  \label{eq:msd_critical_corr}
  \delta r^2(t;\epsilon=0)\simeq A t^{2/z} \left(1+C t^{-y}\right).
\end{equation}

The dynamic correction exponent $y$ can be related to the static correction exponent $\Omega$ via the exponent relation
\begin{equation}
  \label{eq:y_Omega}
  y \dw=\Omega \df\,;
\end{equation}
$\dw$ denotes the walk dimension introduced in \sec{percolation}, and $\df=d-\beta/\nu$ is the fractal space dimension. This relation has been derived by two of us within a cluster-resolved scaling theory for percolation~\cite{PhD_thesis:2006}. The main idea of the derivation is to consider a propagator for tracer particles restricted to clusters of size~$s$. Including an irrelevant parameter of the propagator, similar as in Ref.~\cite{Lorentz_PRL:2006}, generates the leading dynamic correction, \eq{msd_scaling_corr}, as well as the leading correction to the cluster size distribution~$n_s$,
\begin{equation}
n_s \sim  s^{-\tau} \left[1+O(s^{-\Omega})\right] \quad \text{for} \quad s\to\infty,
\end{equation}
with the Fisher exponent $\tau=1+d/\df$. For lattice percolation in $d=3$, the cluster size distribution has been analyzed with extensive Monte Carlo simulations, and the static correction exponent has been determined to $\Omega=0.64\pm 0.02$~\cite{Lorenz:1998}. Assuming that lattice and continuum percolation share the same geometric exponents, one calculates for the three-dimensional Lorentz model $y=0.34$.

For ballistic particles, the corrections are dominant in the critical regime, and they are well described by approximating $\Delta_\pm(\hat t)\approx C$. The corrections to the diffusion coefficient are encoded in $\Delta_-(\hat t)$ for large $\hat t$. They are not small for Brownian particles, and such a simple approximation will fail. But if these deviations are taken into account by hand as above, it seems reasonable to use $\Delta_\pm(\hat t)\approx C$ again. The value of $C$ may be fitted at the critical density by means of \eq{msd_critical_corr}, see inset of \fig{msd-brown-scal}b. Our data corroborate the correction exponent over two decades in time, but the determination of $C$ is hindered due to statistical noise; the data permit a range $9\leq C \leq 15$ (in units $t_0=1$).
Note that the correction amplitude is substantially larger than for ballistic particles, explaining the poor convergence to the long-time asymptote.
Away from the critical density, we have rescaled the mean-square displacements again, taking into account the discussed corrections. Plotting $\delta r^2(t)/A t^{2/z}(1+C t^{-y})$ vs. time rescaled with the measured diffusion coefficients in \fig{msd-brown-scal}b, the data collapse perfectly onto the the scaling function $\widehat{\delta r}{}^2_-(\hat t)$.

The universality hypothesis predicts further that different systems are described by the same scaling functions, provided the latter encode the renormalization flow between the same two fixed points;
in the present example, the fixed points are given by the critical point and the diffusive long-time limit.
In \fig{msd-brown-scal}b, we have also included the collapsed mean-square displacements for ballistic particles from Ref.~\cite{Lorentz_PRL:2006}. An excellent match with the data for Brownian particles is observed, including the crossover regime over more than 5 non-trivial decades in time. Whence the amplitude $A$ and the diffusion coefficients are determined (cf. Figs.~\ref{fig:D_ell_xi3d} and \ref{fig:msd-crit+rect}), no adjustable parameter enters the plot.
One concludes that a single universal scaling function $\widehat{\delta r}{}^2_-(\hat t)$ describes the crossover from anomalous to normal diffusion for both ballistic and Brownian particles.

\begin{figure}
\includegraphics[width=\figwidth]{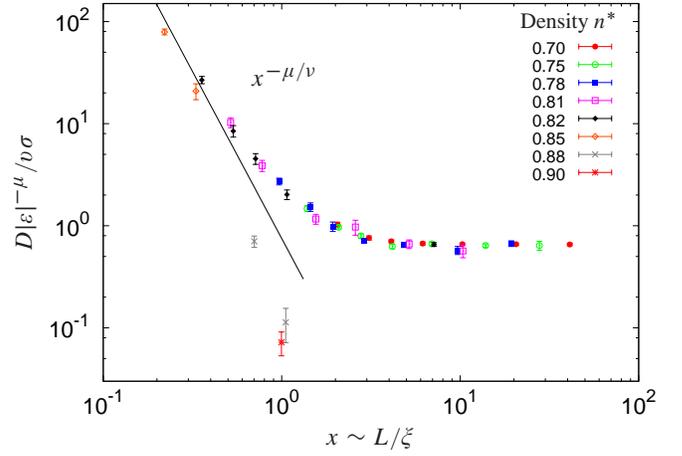}
\caption[Finite-size scaling for diffusion coefficients]{(Color online) The rescaled diffusion
  coefficients (for ballistic particles) collapse on the scaling functions
  $\widehat{D}_\pm(x)$, which describe the finite-size effects. For $L\ll\xi$, the
  rescaled diffusion coefficient diverges asymptotically with exponent
  $\mu/\nu$, see \eq{D_finite_size}.}
\label{fig:D_scaling}
\end{figure}

\subsection{Finite-size scaling}
\label{sec:finite_size}

The size of the simulation box $L$ limits the size of the largest finite
clusters; hence it acts as an upper cutoff on the correlation length.
For very small system, $L\ll\xi$, the correlation length becomes irrelevant, and
the intrinsic (macroscopic) length scale is given by the box size. For very
large systems, $L\gg\xi$, the finiteness of the box can be neglected.
Observables and scaling functions are thus decorated with an additional
parameter, $L$ and $L/\xi$, respectively. We extend
the scaling ansatz for the mean-square displacement, \eq{msd_scaling}, to
\begin{equation}
\delta r^2(t; \epsilon, L)= t^{2/z}\widetilde{\delta r}{}^2(t\ell^{-z},L/\xi),
\end{equation}
and infer for the diffusion coefficient, taking $t\to\infty$,
\begin{multline}
\label{eq:D_finite_size}
D(\epsilon,L) =\xi^{-\mu/\nu}\widehat{D}_\pm(L/\xi) \\
\sim \begin{cases}
L^{-\mu/\nu} & \text{for} \quad L\ll \xi, \\
O\left(e^{-L/\xi}\right) & \text{for} \quad L\gg \xi \quad\text{and}\quad \epsilon\to 0^+, \\
|\epsilon|^\mu & \text{for} \quad L\gg \xi \quad\text{and}\quad \epsilon\to 0^-.\\
\end{cases}
\end{multline}
Note that in small systems, the diffusion coefficient becomes
independent of $\epsilon$ close to the critical density. In particular,
diffusion is not blocked for $n^*>n^*_c$ as long as the box size is smaller than the
correlation length. Diffusing particles exist above $n^*_c$ since the periodic boundary conditions turn large finite clusters that would exceed the box into infinite clusters.
The predicted finite-size scaling of the diffusion coefficient is nicely observed in Fig.~\ref{fig:D_scaling}, where
data for different box sizes and densities above and below the threshold collapse into the two master curves $\widehat{D}_\pm$. It follows that finite-size effects in the Lorentz model may be neglected as soon as $L/\xi\gtrsim 5$, and most importantly, our foregoing analysis is not spoiled by the finiteness of the simulation box.

\begin{figure}
\includegraphics[width=\figwidth]{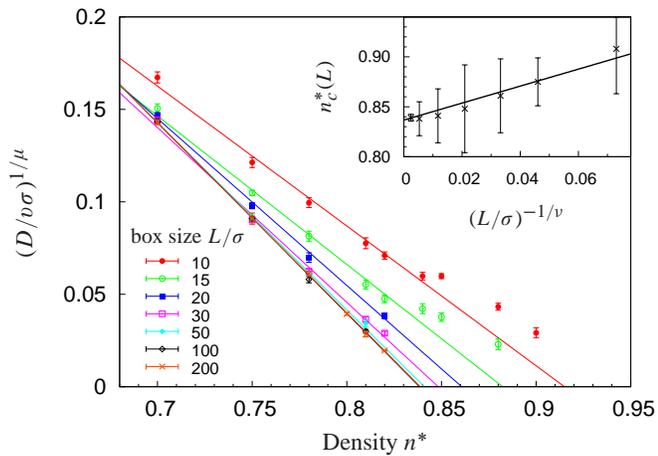}
\caption{(Color online) Rectification plot of the diffusion coefficients of ballistic particles for various box sizes. Straight lines are fits to the data for $n^*\leq
  0.82$, their zeros yield the effective critical densities $n^*_c(L)$. Inset:
  extrapolation of $n^*_c(L)$ to infinite systems.
}
\label{fig:diffusion_rectification_all3d}
\end{figure}

The finite-size effects pretend a shift of the percolation threshold towards the localized regime~\cite{Stauffer:Percolation}.
We have determined the critical density from the axis intercept of the linear extrapolation
of $D(n^*)^{1/\mu}$ vs. $n^*$, see \fig{D_ell_xi3d}c and \fig{diffusion_rectification_all3d} for different box sizes. Rewriting \eq{D_finite_size} as
\begin{equation}
D(\epsilon,L)^{1/\mu}=L^{-1/\nu}\widetilde D(\epsilon L^{1/\nu})
\end{equation}
with an analytic scaling function $\widetilde D$, one infers that $D(n^*)^{1/\mu}$ is shifted with increasing box size and that the effective critical density $n^*(L)$ deviates from the true one as
\begin{equation}
n^*_c(L) -n^*_c \sim L^{-1/\nu}.
\end{equation}
Plotting the effective critical densites $n^*_c(L)$ against $L^{-1/\nu}$ in the inset of \fig{diffusion_rectification_all3d} indeed results in a straight line. The extrapolation to infinite systems allows a more precise determination of the
critical density, we obtain $n^*_c=0.837\pm0.002$. We have checked
that the result is robust against a possible error in the critical
exponent $\nu$ of the order of $0.01$. Therewith, our previous estimate from \sec{simulation_results}, based on a fixed box size, is improved and agrees very well with the result from
\Citet{Rintoul:2000}, $n^*_c=0.8363\pm0.0024$.

\section{Conclusion}
\label{sec:conclusion}

The dynamics of ballistic and Brownian particles was investigated in a heterogeneous environment close to the localization transition. It has been demonstrated that both systems share the same phenomenology on scales where the microscopic details are not resolved anymore. We have further corroborated our previous findings that the localization transition is induced solely by a change in the topology of the medium: the percolation transition of the void space. Most importantly, our data substantiate that both Brownian and ballistic particles in a percolating medium belong to the same dynamic universality class. Taking into account the leading corrections to scaling, we have extracted the universal scaling function for the crossover from anomalous, subdiffusive transport to normal diffusion. The crossover is found to vary remarkably slowly, spanning at least 5 decades in time. In addition, it is an interesting observation that the asymptotic regime is much slower approached by Brownian particles; this statement refers equally well to the asymptotic behavior of the mean-square displacement and to the suppression of the diffusion coefficient upon approaching the localization transition. As a consequence, to observe the genuine asymptotic power law of the anomalous transport is an experimental challenge; very large time windows are required to distinguish universal behavior from transient crossover phenomena.

Although the divergent length scales $\ell$ and $\xi$ cover only a decade in the investigated parameter regimes, we find anomalous transport already over many decades in time. These findings suggest that in experimentally accessible time windows, it is likely to observe apparent power laws with varying exponents, even if the spatial heterogeneities extend just over a decade in length scale. In particular, the Lorentz model provides a generic mechanism for anomalous transport in spatially heterogeneous media.

As mentioned in the introduction, protein transport in cells is anomalous due to the macromolecular crowding. For possible applications to such highly complex systems, the robustness of the presented scenario has to be discussed.
Since the constituents of a cell interact certainly not via a hard potential, it is tempting to replace the obstacles by soft spheres. As long as the kinetic energy of the (ballistic) particle is fixed, the dynamics of a single hard particle, however, can be mapped one-to-one to a soft potential. A canonical ensemble seems more natural for a soft potential, but subsequent averaging with the Boltzmann weight smears out the transition and the critical properties since some fast particles can always overcome the obstacles. We have checked for the two-dimensional Lorentz model that this approach indeed reproduces simulation data for the diffusion coefficient~\cite{Moreno:priv_comm_2004}.

%

\section*{Acknowledgments}

We gratefully acknowledge useful discussions with J.~Cardy and S.~Dietrich on dynamic scaling and corrections to scaling, and with T.~Voigtmann on the simulation of a Brownian particle. We thank R.~Schilling for drawing our attention to Ref.~\cite{Rintoul:2000} and W.~G{\"o}tze for his continuous interest in the topic. F.H.\ acknowledges financial support from IBM Deutschland and the German Excellence Initiative via the program ``Nanosystems Initiative Munich'', and T.M.\ thanks for support from the International Doctorate Program ``Nano-Bio-Technology''. Computing resources were kindly provided by the Leibnizrechenzentrum M\"unchen.

\begin{appendix}
  \section{Short-Time Velocity Correlations}
  \label{sec:short-time-velocity}

  In \sec{brownian-dynamics}, we have pointed on the effect of
  non-vanishing velocity correlations for $t>\tauB$ in the Brownian dynamics
  simulations. Here we explain this behavior quantitatively within a toy model.

  The essential components for the explanation are, first, a random walk which
  is continuous in space but evaluated on a fixed time grid, and second, a hard
  wall restricting the available volume through ballistic reflections. In the
  most simple configuration, we have a point-like particle performing a
  two-dimensional random walk between two parallel walls.


  The calculation of the velocity autocorrelation function $\psi(t)$ is elementary
  for the first two time steps $\tauB$. Choosing a coordinate system with the
  walls parallel to the $y$-axis, we parametrize the velocity during time
  step $n$ by an angle $\varphi_{n}$ to the $x$-axis, $\vec v_{n}(t)=\vec
  v_{n}(n\tauB<t<(n+1)\tauB)=v(\cos\varphi_{n},\sin\varphi_{n})$. There are four
  cases to be distinguished, depending on whether the particle collides with the
  wall or not during the first and the second time step, respectively. As an
  example, when we have collisions in both cases and a fixed absolute value $v$
  of the velocity, the correlation between the initial velocity $\vec v_{0}(0)$
  and the velocity one time step later, $\vec v_{1}(t>\tauB)$, reads
  \begin{equation}
    \langle\vec v_0(0)\dotprod\vec v_1(\tauB+\Delta t)\rangle =
    v^2\overline{(\sin\varphi_0\sin\varphi_1-\cos\varphi_0\cos\varphi_1)}.
  \end{equation}
  By geometric reasoning, one can
  reduce the averages to a double integral to be solved numerically,
\begin{multline}
  \label{eq:vacf_wall}
  \langle\vec v_0(0)\dotprod\vec v_1(\tauB+\Delta t)\rangle=\\
  \frac{2v^3\tauB}{L\pi^2}\int\limits_0^{1}\!dy\int\limits_{\varphi_0}^{\psi_0}\!d\varphi_0'\cos\varphi_0'\sqrt{1-
    \left( \frac{\tauB}{t} \right)^2\! \left( \cos\varphi_0'+y \right)^2},
\end{multline}
where $L$ is the distance between the walls, and the boundaries of the
inner integral are defined by
\begin{equation*}
  \cos\varphi_0=-y; \;
  \cos\psi_0=    
  \begin{cases}
    -\left(
      y+\frac{t}{\tauB}
    \right),& y+\frac{t}{\tauB}<1,\\\cos\pi,& y+\frac{\Delta t}{\tauB}\geq 1.
  \end{cases}
\end{equation*}
The same applies to the other three cases to be considered.

  \begin{figure}
    \centering
    \includegraphics[width=\figwidth]{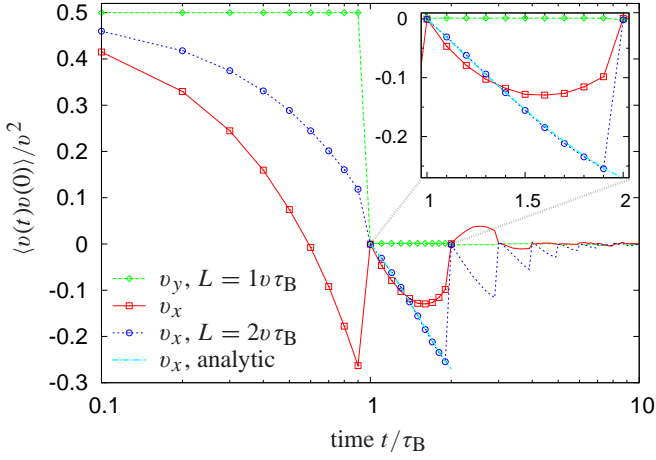}
    \caption{(Color online) Velocity-autocorrelation function for the two-dimensional motion of
      a point-like particle moving between two walls oriented parallel to the
      $y$-axis. The walls are separated by a distance $L$ in $x$-direction. The
      particle is reflected ballistically at the walls, and the direction of the
      velocity is changed randomly in intervals $\tauB$. The inset displays a
      magnification of the regime $\tauB<t<2\tauB$ on the linear scale. The
      analytic curve responding to \eq{vacf_wall} is shown for $L=2v\tauB$; it
      matches perfectly the simulation results.}
    \label{fig:vacf-simple}
  \end{figure}
  
The results obtained from this calculations are in excellent agreement with
simulations of the toy model, see \fig{vacf-simple}. In principle,
$\psi(t)$ can be obtained for longer times $t>2\tauB$ too, but the
calculations are quite involved. \vfill

\end{appendix}

\bibliographystyle{apsrev}
\bibliography{lorentzgas}

\end{document}